\documentclass{ieeetj}
\usepackage{cite}
\usepackage{amsmath,amssymb,amsfonts}
\usepackage{algorithmic}
\usepackage{graphicx,color}
\usepackage{textcomp}
\usepackage{xcolor}
\usepackage{hyperref}
\hypersetup{hidelinks=true}
\usepackage{algorithm,algorithmic}
\def\BibTeX{{\rm B\kern-.05em{\sc i\kern-.025em b}\kern-.08em
    T\kern-.1667em\lower.7ex\hbox{E}\kern-.125emX}}
\AtBeginDocument{\definecolor{tmlcncolor}{cmyk}{0.93,0.59,0.15,0.02}\definecolor{NavyBlue}{RGB}{0,86,125}}

\def\OJlogo{ }
\def\seclogo{ }

\def\authorrefmark#1{\ensuremath{^{\textbf{#1}}}}

\begin{document}

\markboth{}{Westhausen and Meyer}

\title{Binaural multichannel blind speaker separation with a causal low-latency and low-complexity approach}

\author{Nils L. Westhausen\authorrefmark{1}, Student Member, IEEE, Bernd T. Meyer\authorrefmark{1}}
\affil{Communication Acoustics, Carl von Ossietzky University, Oldenburg, Germany}
\corresp{Corresponding author: Nils L. Westhausen (email: nils.westhausen@uni-oldenburg.de).}
\authornote{This work was funded by the Deutsche Forschungsgemeinschaft (DFG, German Research Foundation) under Germany's Excellence Strategy – EXC 2177/1 - Project ID 390895286 and by - Project-ID 352015383 - SFB 1330.}

\begin{abstract}
 In this paper, we introduce a causal low-latency low-complexity approach for binaural multichannel blind speaker separation in noisy reverberant conditions. 
 The model, referred to as Group Communication Binaural Filter and Sum Network (GCBFSnet) predicts complex filters for filter-and-sum beamforming in the time-frequency domain. 
 We apply Group Communication (GC),  
 i.e., latent model variables are split into groups and processed with a shared sequence model with the aim of reducing the complexity of a simple model only containing one convolutional and one recurrent module.
 With GC we are able to reduce the size of the model by up to 83\% and the complexity up to 73\% compared to the model without GC, while mostly retaining performance. 
 Even for the smallest model configuration, GCBFSnet matches the performance of a low-complexity TasNet baseline in most metrics despite the larger size and higher number of required operations of the baseline.
\end{abstract}


\maketitle

\section{Introduction}
Enhancing speech signals in reverberant multi-talker scenarios with additional noise is a challenging problem in the context of assistive listening devices.
One approach for speech enhancement and improving speech perception is speaker separation, which could alleviate challenges that hearing-impaired listeners face in many-talker social interaction.
Speaker separation has seen major breakthroughs in recent years when deep-learning strategies were applied to the problem, both in the time-frequency domain \cite{deep_cluster, upit, gridnet} or by performing time-domain audio separation \cite{ConvTasnet, dprnn, sudormrf}.
However, most of the proposed algorithms are offline approaches, while causal low-latency algorithms are required for hearing aids.
Further, in the context of assistive listening devices, which usually exhibit multiple microphones near the left and right ear, spatial multi-channel information could be exploited for improved source separation. 
A number of multichannel approaches for speech enhancement and speaker separation have been proposed \cite{causal_unet_multi, low_latency, waspa_09_subband, fasnet}.
All of the above-mentioned algorithms are based on architectures with a large computational footprint and are not compatible with small-footprint hardware, e.g., hearing aids.
The majority of these solutions does not consider low-latency constraints \cite{causal_unet_multi, waspa_09_subband} and they are implemented as multiple input single output (MISO) system, while for binaural hearing aids, it would be favorable to use binaural, multiple input multiple output (MIMO) algorithms that could preserve spatial cues.
Such algorithms were also proposed for binaural blind speaker separation \cite{convtasnet_bin} and speech enhancement \cite{marvin_2022}.
However, these approaches use the \emph{causal} ConvTasnet as a backend which has because of their convolutional layers with large receptive fields a large memory footprint originating from the large amount of required rolling buffers. This memory footprint can make the application on memory redistricted low-power devices difficult.

In \cite{groupcomm}, Luo \emph{et al.} refined the sub-band LSTM approach from \cite{waspa_09_subband} to a concept of grouping the latent representation and using a shared sequence model across groups for reducing size and complexity of models.
To share information across parallel groups, a Group Communication module (GC) was proposed.
It was shown that grouping with GC can strongly reduce the size and the complexity of models with only minor performance drops on various architectures \cite{gc3}.

Some of the above-mentioned algorithms have been tested as strategy in hearing-aid processing: 
In \cite{Bramsløw}, Bramsløw et al. showed that single channel neural-network based speaker separation can increase the intelligibility of 2-speaker mixtures when the two separated signals are presented to the right and left ear, respectively.
The benefit was even larger when the listener could choose the desired speaker.

In this study, we use a system that meets many requirements of real-life hearing devices.
We do a in depth evaluation of the \textbf{\underline{B}}inaural \textbf{\underline{F}}ilter and \textbf{\underline{S}}um BFSnet a model adapted to MIMO binaural multichannel blind speaker separation from the Group Communication filter-and-sum network (GCFSnet) introduced in \cite{gcfsnet} for low-latency low-complexity hearing-aid speech enhancement.

The model applies complex-valued filter-and-sum beamforming and directly predicts the complex-valued filter output for separating the talkers.
Additionally, a complex-valued post filter is estimated for each channel individually with the aim of further increasing performance and robustness.
BFSnet provides a low latency since it uses a window length of 2\,ms and a 1\,ms shift. 
It features a simple topology based on a convolutional and a recurrent module and allows for reducing complexity by using grouping and Group Communication \cite{groupcomm} (GC) with transform average concatenate (TAC) \cite{gc3}. The model using GC is referred to as GCBFSnet.

We are interested in the trade-off between model performance, number of parallel groups for the grouped parts of the model, and model size controlled with the hidden size of the groups. 
Four objective output metrics are reported that either focus on single-channel improvements (scale-invariant signal-to-distortion ratio, SI-SDR, and perceptual evaluation of speech quality, PESQ) or on binaural/better-ear perception, i.e., the modified binaural short-time intelligibility (MBSTOI) model and the hearing-aid speech perception index (HASPI).

We are comparing the GCBFSnet to three baseline models based on the common causal ConvTasNet \cite{ConvTasnet}. The first is the binaural ConvTasNet as described in \cite{convtasnet_bin}. For the second model this approach is adapted to full MIMO processing for fair comparison. For the third the MIMO model is used with additional GC to create a low-complexity baseline.

The models are trained and used for speaker separation in a scenario with two speakers in a reverberant environment with diffuse noise. 
This scenario is static and well-defined, but also quite challenging due to the task, noise, and reverberation. It also represents a common communication situation which can be analyzed with respect to spatial separation of speakers and reverberation time (as presented in a post analysis of results). 
For creating the training data we used the Libri 2mix corpus \cite{cosentino2020librimix} and a large amount of simulated multichannel binaural room impulse responses (BRIR) to represent a large variety of acoustic conditions and source-and-receiver positions.

\section{Methods}

\subsection{Acoustic scenario}
\begin{figure}[t]
  \centering
  \includegraphics[width=1.00\linewidth]{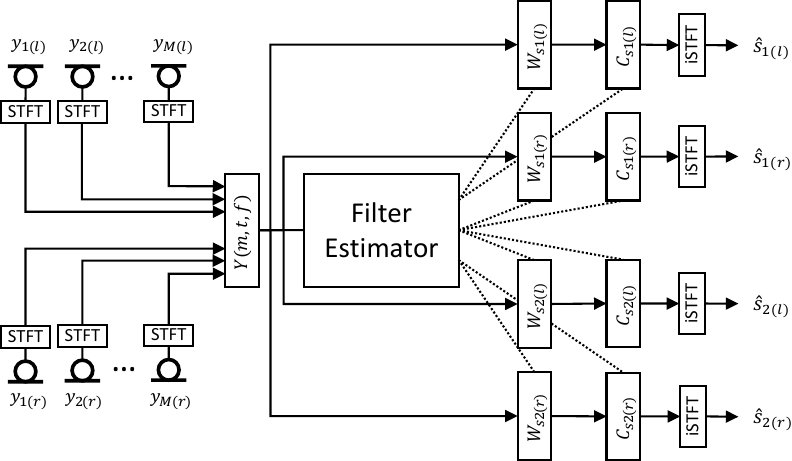}
  \caption{Illustration of binaural spatial filtering structure with post filter for separation of two speakers with $M_l$ microphones on the left and $M_r$ microphones on the right.}
\label{fig:filter-structure}
\end{figure}

For this study, we are considering scenarios with two speakers and a noise source. 
The proposed model estimates filter coefficients in the frequency domain. We therefore first apply the multichannel short-time Fourier transform (STFT), which can be written as
\begin{equation}
    Y(m, t, f) = X_{1}(m,t,f) + X_{2}(m,t,f) + X_{N}(m,t,f)
\end{equation}
where $m$, $t$, $f$ are the microphone, frame and frequency index, respectively. $X_{i}$ corresponds to the complex time frequency (TF) representation of the anechoic speech signal $s_{i}(n)$ of speaker $i$ convolved with the corresponding binaural multichannel room impulse response $h_{i}(m,n)$. $Y$ is the mixture while $X_{N}$ is the TF-representation of the noise $d(n)$ convolved with the BRIR $h_d(m,n)$.
Our goal is to extract the binaural speech signals $s_{i(l,r)}(n)$ only containing the direct part of the reverberant speech.
The estimation is performed by complex filter-and-sum beamforming for the left $l$ and right $r$ channel equally as
\begin{equation}
\label{eqn:filter_channel}
{\Tilde{S}}_{i(l,r)}(t, f) = \sum_{m=1}^{2M} Y(m, t, f) \cdot W_{i(l,r)}(m, t, f)
\end{equation}
where $\Tilde{S}_{i(l,r)}$ denotes the intermediate estimated binaural TF-representation of speaker $i$ and $W_{i(l,r)} \in \{ z \in \mathbb{C} \mid -1 \leq \Re(z) \leq 1, -1 \leq \Im(z) \leq 1 \}$ are the filters estimated by the model. $M$ is the number of microphones on each side.
Additionally, a single channel post filter is applied to the left and the right channel of each source separately 
\begin{equation}
\label{eqn:filter_post}
\hat{S}_{i(l,r)}(t, f) =  \Tilde{S}_{i(l,r)}(t, f) \cdot C_{i(l,r)}(t,f).
\end{equation}
$C_{i(l,r)} \in \{ z \in \mathbb{C} \mid -1 \leq \Re(z) \leq 1, -1 \leq \Im(z) \leq 1 \}$ denotes the complex filter for the left and right channel of speaker $i$. 
$\hat{S}_{i(l,r)}(t, f)$ is transformed back to the time-domain by an iSTFT.
The general algorithm structure is illustrated in \autoref{fig:filter-structure}.

\subsection{Architecture}
\begin{figure*}[t]
  \centering
  \includegraphics[width=1.00\linewidth]{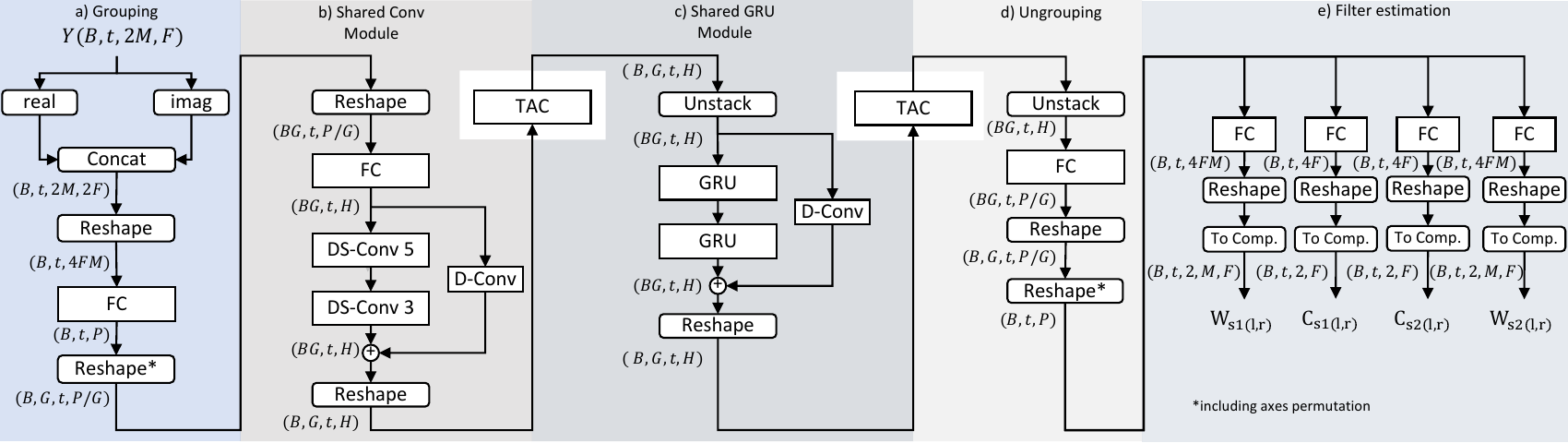}
  \caption{Illustration of the proposed filter estimation model. }
\label{fig:filter-network}
\end{figure*}
The proposed architecture (visualized in \autoref{fig:filter-network}) has five modules. (a) A grouping module combines the features and splits the latent representation in equally sized groups. (b) The Conv module is included since a convolutional layer can extract local temporal information, which could be especially important when a very short frame length is chosen as in our approach. (c) Due to its recurrent connections, the GRU module should account for long-term temporal dependencies. (d) An ungrouping module recombines the outputs of the groups to one latent representation from which the filters are predicted by the (e) filter estimation module. 

The concatenated real and imaginary part of $Y$ serve as features to the grouping module; the multi-microphone representations are reshaped into a single input vector. This vector is linearly projected by a fully-connected layer (FC) of size $(4FM) \times P$. 
The projection is split into equal groups of size $P/G$ where $G$ is the number of parallel groups. This is inspired by the grouping performed in \cite{groupcomm}. 
Grouping is a from of weight sharing and implies the reuse of a sequence model for equal splits of a latent dimensions. 
Hence, the amount of trainable weights can be reduced while retaining a similar performance when keeping
the amount of computations constant; when using a larger number of groups, performance could also increase compared to the system without GC. 

Next, the groups are processed by a Conv module with shared weights across all groups.
The Conv module contains an FC layer of size $(P/G) \times H$ and parametric rectified linear unit (PReLU) activation. It is followed by causal depth-wise separable convolution (DS-Conv) with kernel size 5 across time followed by a DS-Conv with kernel size 3, both with PReLU activation as well. An additive skip connection with a depth-wise convolution (D-Conv) and kernel size 1 is used in a branch parallel to the DS-Conv operations.
This kind of skip-connection is inspired by \cite{deepfilternet2}. D-Conv and DS-Conv were chosen to reduce the number of weights and complexity compared to their full Conv layer counterparts.
For frame-by-frame implementation of the Conv module, two rolling buffers are required for saving the previous inputs to the D-Conv operations.

TAC \cite{fasnet_tac} is utilized for GC \cite{groupcomm} between the Conv and the following module, which is motivated by findings from Luo et al. who reported superior results in comparison to using BLSTMs \cite{gc3}. Further, TAC requires fewer operations than BLSTM processing and can be parallelized during inference.
TAC consists of three FC layers with PReLU activation. The first FC layer is shared between all groups and transforms the latent presentation of the groups of size $H$ to size $2H$. Next, the average over the group axis is calculated. The average is processed by the the second FC layer with size $2H \times 2H$. This output is concatenated to the transformed output of each group created by the first FC layer and mapped by the third FC layer from $4H$ to size $H$. In a last step, the input to TAC is added to the output as an additive skip connection.    
Compared to \cite{fasnet_tac} and \cite{gc3} we use TAC with a hidden dimension of $2H$ instead of $3H$ to reduce the complexity. Preliminary experiments have not shown a large degradation from the reduction of the hidden dimension.

The next module is a shared GRU module containing two stacked GRUs with $H$ units with an additive D-Conv skip connection. 
The GRU-Module is followed again by TAC.
The combination of first using convolutional layers followed by recurrent layers have shown to be beneficial in previous studies \cite{google_convrnn, Valin2020}.

In the following Ungrouping module, a shared FC layer linearly projects the latent representation of the groups of size $H$ to size $P/G$. 
The projected latent representations of the groups are concatenated to form a latent representation of size $P$. From this representation, the real and complex part of the filters $W$ and $C$ are predicted by FC layers of size $P \times (4FM)$ for $W$ and size $P \times (4F)$ for $C$ with a tanh activation in the filter estimation module. The outputs of the FC layers are reshaped and combined to the final complex filters.
\subsection{Loss function}
The compressed spectral MSE (cMSE) previously used for multichannel sound source separation \cite{db_net_braun} is used as loss function. The cMSE is defined as follows
\begin{equation}
    L_{\mathrm{cMSE}} = (1-\alpha) ||\hat{X}|^c - |X|^c|^2 + \alpha |\hat{X}^c - X^c|^2,
\end{equation}
where $X$ and $\hat{X}$ denote true and estimated STFT of the binaural target signal, respectively. The sum, the frequency, time and channel indices are omitted for brevity. 

Note that STFT parameters used for the cost function 
does not have to be identical to the STFT for the filtering framework.
$X^c = |X|^c \frac{X}{|X|}$ denotes the magnitude compression of $X$ with $c$. $\alpha$ is a weight factor to balance the MSE between complex-valued spectra and magnitude spectra. 
This loss function was chosen since the power-law compression is beneficial during training. It reduces the dominance of large values.
$c$ and $\alpha$ were set to 0.3 for all experiments. These values were found to be a good trade-off between signal quality and interference reduction \cite{braun_loss}.
The loss is combined with utterance-wise permutation-invariant training \cite{upit} (uPIT ) to address the permutation problem of speaker separation. When applying uPIT the left and right channel of each speaker is seen as a fixed set to enforce a fixed channel order and so not allowing random switching of the left and right channel.

\subsection{Datasets and data generation}
As basis for our training, evaluation and test set, we created a spatial version of the LibriMix Corpus \cite{cosentino2020librimix}. 
LibriMix includes the noise from the WHAM! corpus \cite{Wichern2019WHAM}, which was gathered in cafes, bars and public spaces. 
The LibriMix corpus has a large amount of data and a large variety and number of speakers which is beneficial for training robust models.

To create spatial signals, we simulated approximately 60k rooms with RAZR \cite{wendt2014a} with RT60 values drawn equally from the range of 0.1 to 1~s. The rooms have ceiling heights equally drawn from 2.5 to 4.5~m, widths from 3 to 10~m and surface areas between 12 and 100~$\mathrm{m}^2$. 
As a basis for the spatial rendering we are using the high-resolution HRTF set from \cite{Thiemann2019} of the B\&K Hats with behind-the-ear hearing-aid dummies with three microphones on each side (front, mid and rear). In total, 6 hearing-aid channels are available.

For each room, three point sound sources are modeled at randomly drawn positions.
The first two sources are used for speech signals while the third is used for the noise source. The speech sources are at least 1~m away from walls and positioned in the range between 0.75 to 2~m away from the receiver at a random angle at a height between 0.9 to 1.8~m. The noise source has the same settings except its distance to the receiver, which is at least 1~m with no upper bound. The receiver is positioned at random up to 1~m from the center of the room at a height between 0.9 to 1.8~m.
From RAZR, we can obtain either the full BRIR $H$ or segments that relate to the direct component or the impulse response $H_{direct}$ or the early reflections and late diffuse reverberation ($H_{early}$ and $H_{late}$, respectively). 
 This is beneficial since it enables us to use $H_{direct}$ as a properly delayed training target to also include dereverberation and have a clean target for the objective measures. 

The training data comprises the 50.8k utterances and noise files of the min train-360 split from LibriMix. 
Training is validated with the 3k mixtures from the min dev split, while testing is performed with the 3k mixtures from the max test split. 
During training, the training epochs are created online to have a larger variety compared to a fixed training set (since similar but not identical signals are used in different epochs). For each training sample random speakers and noise utterances are chosen in a first step. A random, 4-second long signal is extracted from the utterances.
Signals are convolved with BRIRs from a random room where the speech signals for the mixture are convolved with the full BRIR $H$ while the speech signals for the target are convolved with $H_{direct}$. 
The noise utterance is convolved with $H_{late}$ of the interferer BRIR, which is the output of a feedback delay network, to create a diffuse noise. 
We found it to be convincing virtual representation of the WHAM! noise. 
Next, the utterances of the second talker and the noise utterance are normalized by their corresponding speech-weighted better-ear SNRs on the front microphones relative to the first talker as it was done in \cite{graetzer21_interspeech}. The second talker's signal is scaled by a gain drawn from a normal distribution $\mathcal{N}(0, 4.1^2)$ dB and the noise is scaled by random value from the normal distribution $\mathcal{N}(6.2, 4.4^2)$.
These distributions correspond to the original distribution of LibriMix data.
In a final step, the data is mixed and scaled to a random value drawn from $\mathcal{N}(-26, 5^2)$ dB~FS (relative to full scale) to simulate recording level variability. All other signals are scaled correspondingly. 

The process for creating in-training validation data is similar, but the fixed combinations of talkers and noise and the SNR of the original material is used. For all signals, the first 4s are chosen instead of random parts. 

The test set is created offline from the LibriMix max test split and uses the whole utterances as well as the original SNRs and the original input scales.

Training, validation and test datasets have their unique set of BRIRs and utterances. There is no overlap between the splits.

\subsection{Training configuration}
\label{subsec:train_conf}
The model is trained for 100 epochs with a batchsize $B$ of 32. ADAM is used as optimizer with an initial learning rate of 1e-3. The learning rate is multiplied by 0.98 every two epochs. If the loss on the validation set does not decrease for 5 consecutive epochs, the learning rate is multiplied by 0.8. 
For gradient clipping, we use AutoClip \cite{autoclip} with $p=10$ for smoother training and better generalization.
The models are trained on a workstation with 4 Nvidia RTX A5000 and 32 CPU cores. Each training of a model uses only one GPU. The training setup uses PyTorch 2.0.1.
The frame length of the filtering framework is 2\,ms with a shift of 1\,ms and an FFT-length of 32. For STFT and iSTFT calculation, a $\sqrt{Hann}$ window is applied. The STFT for the loss function is set to 20\,ms window length and 10\,ms shift with an FFT-length of 320.
$P$ is set to 256 for all configurations. When $G$ equals 1, TAC is deactivated.
The front and the rear microphone from each side is used as input for all models, where the direct part of the signals of the front microphone is used as reference. 
\begin{table*}
\caption{Results for the noisy reverberant test set in terms of evaluation metrics. 
Model complexity is reported in terms of parallel groups $G$, numbers of hidden units $H$, model size and multiply-accumulate (MAC) operations. 
The latter are reported for processing 1~s of audio with a frame shift of 1~ms.}
  \centering
  \begin{tabular}{l @{\hspace{0.2cm}}c @{\hspace{0.2cm}}c c c c c c c  } 
    \hline
    \textbf{Model}  & \textbf{$G$}  & \textbf{$H$}   & \textbf{\# Size} $\downarrow$& \textbf{\# MACs}  $\downarrow$& \textbf{HASPI} $\uparrow$& \textbf{MBSTOI} $\uparrow$ & \textbf{Si-SDR} $\uparrow$& \textbf{PESQ} $\uparrow$\\
    \hline
     Unprocessed &  &  & &  & 0.25 & 0.46 & -9.16 & 1.08 \\
     \hline
     BiTasNet &  & 256 & 1.65M (129\%) & 3.43G (268\%) & 0.51 & 0.59 & -1.26 & 1.19   \\
     MiMoBiTasNet &  & 256 & 1.71M (135\%) & 1.84G (145\%) & 0.59 & 0.66 & 0.76 & 1.25    \\
     GCMiMoBiTasNet & 8 & 32 & 242.7K (19\%) & 0.95G (75\%) & 0.52 & 0.59  & -1.65 & 1.19    \\
      \hline
     BFSnet & 1 & 256 & 1.27M (100\%) & 1.27G (100\%) & 0.67 & 0.60 & -0.96 & 1.26   \\
     \hline
     BFSnet & 1 & 128 & 508.7K (40\%) & 0.51G (40\%) & 0.62 & 0.58 & -1.50 & 1.25  \\
     GCBFSnet & 2 & 128 & 804.8K (63\%) & 1.27G (100\%) & 0.69 & 0.61 & -0.65 & 1.27  \\
     GCBFSnet & 4 & 128 & 788.3K (62\%) & 2.14G (169\%) & 0.74 & 0.63 & -0.26 & 1.29   \\
      \hline
     GCBFSnet & 4 & 64  & 359.9K (28\%)& 0.71G (56\%) & 0.68 & 0.61  & -0.82 & 1.27   \\
     GCBFSnet & 8 & 64  & 355.7K (28\%)& 1.15G (91\%) & 0.70 & 0.61 & -0.57 & 1.28   \\
      \hline
     GCBFSnet & 8 & 32  & 248.0K (20\%) & 0.46G (36\%) & 0.66 & 0.59 & -1.20 & 1.26  \\
     GCBFSnet & 16 & 32 & 246.9K (19\%) & 0.68G (54\%) & 0.69 & 0.60 & -1.05 & 1.27   \\
      \hline
     GCBFSnet & 16 & 16 & 219.7K (17\%) & 0.34G (27\%) & 0.64 & 0.58 & -1.72 & 1.25   \\
     GCBFSnet & 32 & 16 & 219.4K (17\%) & 0.46G (36\%) & 0.67 & 0.59 & -1.49 & 1.26   \\
     
    \hline
  \end{tabular}
  \label{tab:results}
\end{table*}
\subsection{Baselines}
Three different configurations of ConvTasnet \cite{ConvTasnet} serve as baseline. The first version is the original causal binaural ConvTasNet (BiTasNet) proposed in \cite{convtasnet_bin}. BiTasNet is a MISO approach, so the model is used twice for predicting the left and the right output channel from a two channel input (left and right). 
It uses parallel encoders and a mask-and-sum mechanism in a learned feature domain which is similar to filter-and-sum processing performed by the GCBFSnet in the time-frequency domain. As in \cite{convtasnet_bin}, 64 filters are used in the linear encoder with a frame size of 2\,ms and a frame shift of 1\,ms. Four stacks of eight 1-D convolutional blocks with causal cumulative layer normalization are utilized for the temporal convolutional network (TCN). This results in 1.65M parameters.

For the second baseline, we adapted this system to a full MIMO (BiTasNetMiMo) system accepting 4-channel input and predicting all output channels directly (so it is not required to run twice) while using the parallel encoders and the mask-and-sum mechanism for all four input channels. 
The TCN of this baseline uses the same configuration as the previous one.

To also include a low complexity baseline, we further adapted the sequence model of the BiTasNetMiMo to include GC with TAC (GCBiTasNetMiMo). We use eight groups with a hidden size of 32 which results in around 243k parameters. For TAC, a hidden size of 64 is used. All models are trained for 100 epochs with the SNR loss as suggested in \cite{convtasnet_bin}, a batchsize of 8 and a learning rate of 5e-4. The learning rate scheduling is performed as explained in \autoref{subsec:train_conf}.
 
\subsection{Evaluation metrics}
The first metric we are using for evaluation is the Scale-Invariant Source-to-Distortion Ratio (SI-SDR) \cite{SI-SDR}, a common measure for source separation quality. 
The second metric is the Perceptual Evaluation of Speech Quality (PESQ) \cite{pesq}.
For SI-SDR and PESQ, the mean over the the binaural signal is calculated. 
To better account for spatial hearing or speech perception in the context of hearing aids, we use the Hearing Aid Speech Perception Index (HASPI) \cite{HASPI} (Range: 0 to 1) as third metric. 
HASPI is used with a better-ear mechanism, i.e., 
its output calculated for both ears and the maximum of these values is returned. 
This is inspired by findings that large amount of speech reception performance in multi-talker scenarios of human subjects can be explained with perception of the ear with the better/higher signal to noise/interferer ratio \cite{edmonds2006spatial_better_ear}.
The last metric is the Modified Binaural Short Time Objective Intelligibility (MBSTOI) \cite{MBSTOI}, which takes binaural processing into account. Its output values range from 0 to 1 (higher is better). 
MBSTOI was used as evaluation metric of the first round of the 1st Clarity Enhancement Challenge \cite{graetzer21_interspeech} while HASPI was used during the first round of the 2nd Enhancement Challenge. For HASPI, the hearing thresholds are set to 0~dB HL (normal hearing configuration) and the the mixture is scaled to represent a playback level of 65~dB SPL (sound pressure level). We use a normal hearing configuration since we are not applying any hearing loss compensation which would be required when applying a hearing loss.
The reference for all metrics is the speech signal convolved with the direct part of the BRIR.

We report the computational complexity of the model in multiply-accumulate operations per second (MACs) and the size of the model in terms of number of parameters.

\section{Results}
The results in terms of objective metrics and the corresponding size and complexity of the models are shown in \autoref{tab:results}.

\textbf{Overall performance}
For all metrics, performance and complexity are positively correlated, while this relation is less pronounced for model size and performance (i.e., a decrease in size does not always result in decreased performance). The model with $G = 4$ and $H = 128$ reaches the best results; compared to BFSnet, about twice the number of MACs are required while the number of weights is decrased by 37\%. 
The lowest average performance with a GCBFSnet model is obtained with $G = 16$ and $H = 16$. The performance increases when keeping the hidden dimension constant and increasing the number of groups. The effect is most prominent when comparing the configuration with $H = 128$. The improvement in terms of HASPI from $G = 1$ (no grouping and TAC) to $G = 4$ is 0.12 and in terms of SI-SDR 1.24 dB while the number of parameters only increases by 280k weights and the complexity increases by a factor of four.
The highest performance of the TasNet versions is reached by MiMoBiTasNet. It produces the highest overall SI-SDR score with 0.76, which is a total improvement of 9.92. 
The original BiTasNet performs worse than MiMoBiTasNet while using twice the amount of required operations. 
GCMiMoBiTasNet matches the performance of BiTasNet except for the SI-SDR while only using a quarter of required MACs. 
Compared to MiMoBiTasNet, the performance in terms of metrics is reduced. 
The grouping mechanism in GCMiMoBiTasNet reduces the amount weights by a factor of 7 and the required number of MACs by a factor of 2 compared to MiMoBiTasNet. 
BiTasNet and GCMiMoBiTasNet are outperformed by all configurations of the (GC)BFSnet in terms of HASPI and PESQ  and also matched or outperformed for SI-SDR and MBSTOI (with the exception of the configuration with $G = 16,H = 16$ and $G = 1,H = 128$).
In terms of HASPI, MiMoBiTasNet is outperfomed by all (GC)BFSnet configurations while for PESQ it is matched by (GC)BFSnet with $G = 16,H = 16$ and $G = 1,H = 128$ and slightly outperformed by all other configurations. 
MiMoBiTasNet produces the best overall scores in terms of MBSTOI and SI-SDR; it is also larger by a factor of eight (no. of parameters) or four (MACs) compared to the smallest GCBFSnet, which produces better HASPI and PESQ scores.

\begin{figure*}
  \centering
  \includegraphics[width=1.00\linewidth]{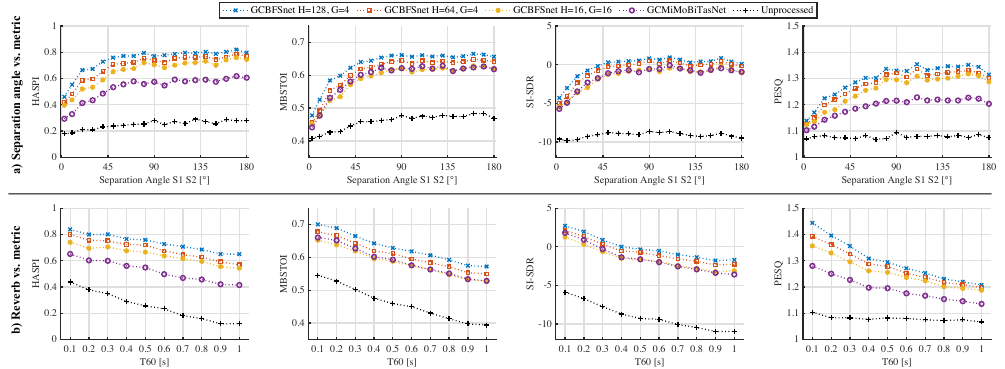}
  \caption{Results in terms of objective metrics plotted versus the separation angle between two speakers (a) and objective metrics plotted versus the reverberation time T60 (s) (b) for the three configurations of the GCBFSnet (representing different levels of complexity: high H=128, G=4, medium H=64, G=4 and low H=16, G=16) and the GCMiMoTasNet as reference.}
\label{fig:results}
\end{figure*}

\textbf{Results relative to the separation angle of the speech sources:}
In the upper row (a) of \autoref{fig:results}, the metrics are plotted versus the separation angle between speaker 1 and speaker 2. The results of the test set are clustered for 10° steps. 
The metrics for the unprocessed signals are relatively constant over the separation angles. Only for MBSTOI a slight constant reduction is visible from 90° to 0°.
The results for all models and metrics following the same trend, the improvement above the unprocessed increases with increasing separation angle up to 90° and saturates above 90°. No large improvements are visible for any of the metrics or models above 90°.
The highest results over all angles and metrics is reached by the GCBFSnet with $G = 4,H = 128$. GCBFSnet with $G = 4,H = 64$ shows a lower performance compared to $G = 4,H = 128$ and $G = 16,H = 16$ shows again slightly lower improvements compared to $G = 4,H = 128$. 
GCMiMoBiTasNet matches the performance of GCBFSnet wit $G = 16,H = 16$ for MBSTOI and SI-SDR while it is clearly outperformed by all visible configurations for HASPI and PESQ.
These results are inline with the overall results as shown in \autoref{tab:results}.

\textbf{Results relative to reverberation time:}
In the lower row (b) of \autoref{fig:results}, the metrics are plotted versus the reverbertion time T60 of the simulated rooms of the test set. The results of the test set are clustered for 0.1s steps from 0.1 to 1~s.
The metrics of the unprocessed signals decrease with increasing T60 for all metrics exept PESQ which stays relatively constant for all T60 values. The overall values of the metrics decrease for all models with increasing T60. For HASPI, MBSTOI and PESQ the improvement above the unprocessed slightly increases with increasing T60 for all models. For PESQ the improvement decreases with increasing T60 for all evaluated models. The order of the ranking of the models is inline with results visible in \autoref{tab:results}.

\begin{table}[!th]   
\caption{Results of the ablation study on for GCBFSnet configurations with different complexity (high: H=128, G=4; medium:H=64, G=4; low:H=16, G=16). For all metrics higher is better.}
  \centering
  \begin{tabular}{l @{\hspace{0.2cm}}c @{\hspace{0.2cm}}c @{\hspace{0.22cm}}c @{\hspace{0.22cm}}c @{\hspace{0.22cm}}c @{\hspace{0.22cm}}c @{\hspace{0.22cm}}c   } 
    \hline
    \textbf{Model}  & \textbf{$W$}  & \textbf{$C$}  & \textbf{$scaling$}  & \textbf{HASPI} & \textbf{MBSTOI} & \textbf{Si-SDR} & \textbf{PESQ} \\
    \hline

     high & \checkmark & \checkmark &  & \textbf{0.74} & \textbf{0.63} & \textbf{-0.26} & \textbf{1.29}   \\
     & \checkmark &  &  & 0.69 & 0.62 & -0.4 & 1.28 \\
    & \checkmark &  & \checkmark & 0.70 & 0.62 & -0.34 & 1.28 \\
      \hline
     medium & \checkmark & \checkmark &  & \textbf{0.68} & \textbf{0.61}  & \textbf{-0.82} & \textbf{1.27}   \\
     & \checkmark &  &  & 0.64 & 0.60 & -1.05 & 1.26 \\
     & \checkmark &  & \checkmark & 0.64 & 0.60 & -0.93 & 1.26 \\
      \hline
     low & \checkmark & \checkmark & & \textbf{0.64} & \textbf{0.58} & \textbf{-1.72} & \textbf{1.25}   \\
     & \checkmark &  &  & 0.62 & 0.57 & -2.03 & 1.23 \\
     & \checkmark &  & \checkmark & 0.61 & 0.57 & -1.97 & 1.24 \\
    \hline
  \end{tabular}
  \label{tab:results_ablation}
\end{table}
\textbf{Ablation study regarding post filter:}
To evaluate the effect of the post filter, an ablation study is conducted. 
Models with different degrees of complexity (high: $G=4,H=128$; medium: $G=4,H=64$, low: $G=16,H=16$) are trained without the post filter while keeping all other parameters constant. 
Since the learned filter values (real and imaginary part) of $W$ and $C$ are each restricted to the interval [-1,+1], $W$ alone can not apply the same maximum amplification as $W$ and $C$ combined. 
For this reason, we also train the same three configuration while applying a 
$scaling$ factor to $W$.
This scaling factor is set $\sqrt{2}$, which compensates for the possible magnitude amplification when $C$ is additionally applied.
The results of this analysis are shown in \autoref{tab:results_ablation}.

For all three configurations, the removal of the post filter results in slight reduction in performance. 
This is also true when a scaling factor is applied to compensate for the potentially decreased magnitude amplification when omitting the post filter. 
Absolute differences for MBSTOI and PESQ are only around 0.01, while slightly larger differences are observed for HASPI and SI-SDR. 
The effects of the post filter appear to be small, yet, they are consistent for all three size configurations. 

\section{Discussion}
\textbf{Effect of model size and grouping:}
The gradual decline in performance with smaller models is consistent with \cite{groupcomm}. Increasing the number of groups while keeping $H$ constant always increases the performance. In general it can be concluded that the performance of GCBFSnet depends more on the computational complexity than on the model size. When keeping the number of MACs constant it can be beneficial to choose a model with grouping over a model without grouping. 
This is indicated by the slightly better performance of GCBFSnet with $H=128,G=2$ compared to the BFSnet with $H=256$, both with 1.27G MACs.
We did not explore smaller numbers of weights in the current study since the number of weights for the filter module is fixed at 175K weights and does not vary with $G$ and $H$. For $G = 16$ and $H = 16$, 80\% of the model parameters are already attributed to the filter module. On the other hand, the sequence part has only 44k weights in this configuration which is a reduction by a factor of 25 compared to the sequence part of BFSnet.
The configuration with $G = 4, H = 64$ appears to be a good trade-off between complexity, size and performance. Generally, a model configuration that fits hardware constraints of a desired platform could be chosen without sacrificing too much performance.
With the decrease in model parameters and computational complexity reported here, the smallest separation model should already be compatible with next-gen hearing-aid system-on-chip (SOC) such as \cite{smartHeap} if weights and operations are quantized to 8-bit integers.

\textbf{Effect of spatial between-speaker angle:}
With increasing azimuth angle between the two speakers, separation performance also increases, which is true for all four objective metrics explored in our paper.  
This is not surprising since a larger angle implies an improved signal-to-interferer ratio at at least one microphone. The model should learn to exploit information from the microphone that is closer to the desired speaker (similar to better-ear-listening in human listeners).
Additionally, due to the joint learning that uses \emph{all} microphone channels as input, the model can implicitly learn to exploit additional cues (such as interaural time and level differences) that are relevant in human binaural unmasking. In future research, it would be interesting to disentangle these two effects.

\textbf{Effect of amount of reverberation:}
The effect of reverberation was analyzed (lower row in Fig.\,3, which shows the absolute output of objective measures). Again unsurprisingly, all metrics degrade as the reverberation time increases. The PESQ score for the unprocessed signals is an exception, which could be caused by a flooring effect: Unprocessed scores are relatively low in the first place, and the lower bound of the PESQ scale is 1 (since it was designed for modeling mean opinion scores from 1 to 5). While the trends of all objective metrics are consistent, we find the HASPI score to be especially interesting since its model output should directly relate to intelligibility (i.e., a mapping function from model output to intelligibility is not required). From this we can estimate that the best-performing system (GCBFSNet, H=128, G=4) increases intelligibility from 44 to 84\,\% compared to the unprocessed baseline at a T60 of 100\,ms. At the same time, the detrimental effect of reverberation can be quantified, since the performance of the best system decreases
to 65\,\% (T60=1000\,ms) which amounts to a relative degradation of 19\% in terms of intelligibility. 
In daily life, we usually communicate in environments with lower T60 values than 1~s, still, the results indicate that reverberation has a large effect which needs to be considered in future speech-processing systems.

\textbf{Effect of the post filter:}
The results in \autoref{tab:results_ablation} show that the post filter can create a slight improvement compared to using only spatial filters. 
When applying a scaling factor to the  spatial filter $W$ (to compensate for amplification of the post filter), the model should be able to learn the same characteristics for the spatial filter as for the combination of the spatial filters and post filter.
However, results indicate a small benefit for learning separate weights over scaling alone.
One factor to explain this is the slightly larger filter prediction module when using the post filters. 
If the increased size of the filtering module would be the only cause, the difference between configurations with and without post filter should be more pronounced for model sizes dominated by the size of the post filter. 
However, the observed benefits are relatively stable across the model configurations shown in \autoref{tab:results_ablation}. 
We therefore assume that the separation into spatial filters and post filter brings some slight benefit for the training process exceeding the mere effect of an increased size of the filtering module, which should be explored in the future.

\textbf{BiTasNet and comparison to (GC)BFSnet:}
 The first baseline system BiTasNet exhibits a relatively low performance but requires lots of MAC operations, which follows from the setup as MISO model, i.e., it has to run twice using two channels of all available four channels. 
 BiTasNetMiMo profits from the four-channel input and shows superior performance to the original BiTasNet. 
 We assume that the reason for the strong performance for MBSTOI and SI-SDR is the loss function, i.e., a time-domain negative SNR function, which optimizes the time domain representation in a beneficial way for these two metrics. 
 
 The reduction in terms of performance metrics from grouping in GCBiTasNetMiMo ($G = 8, H = 32$) relative to BiTasNetMiMo is higher than for BFSnet and GCBFSnet ($G = 8, H = 32$) while still having relatively large computational footprint. 
 The organization of the grouping could be a potential explanation: In GCBFSnet the output of \emph{all} groups is used for filter estimation module, while in GCBiTasNetMiMo the output of the groups
 is directly used as masks (as originally implemented for \cite{gc3}) without a dedicated filter estimation module for utilizing and mixing the results of all groups for all masks.  
While the cost in terms of size and complexity of using the information of all groups for estimating the filters is relatively high because of the larger matrix multiplications, it seems to be beneficial compared to the approach used in GCBiTasNetMiMo. 
 
 Another important difference of the approaches is that (GC)BFSnet predicts filter in the time-frequency domain while BiTasNet approaches are predicting masks in a learned feature domain. 
 Most state-of-the-art source separation or low-latency approaches are waveform-to-waveform \cite{sudormrf} or use a learned feature transform \cite{dprnn}. 
 However, the good average performance of (GC)BFSnet indicates that for low-latency approaches time-frequency domain approaches are still an option. 
This is consistent with recent findings for the related task of speech enhancement
in \cite{low_latency, low_complexity_low_latency} using models with a relatively high complexity compared to (GC)BFSnet.

\textbf{Training and evaluation data:}
To reproduce results from research using speech technology, the use of open-source datasets provided by the community should be preferred over using proprietary data. For the current study, we still decided to create a new dataset, however based on open sources, so the dataset can be made available upon publication. 
The requirements for the dataset were that it should cover scenarios with two speakers and diffuse noise in arbitrary rooms, i.e., the training data should contain a large variety of reverberation times and speakers compared to existing datasets. 
The data from the Clarity Challenges \cite{graetzer21_interspeech} approximate these requirements, but the signals do not cover a mixture of noise and two speakers, as required for separation (first challenge round) or the number of speakers was variable second challenge round. Further, the number of acoustic scenes was limited to a total of 10k scenes for training, validation and test.
In the current study, we included larger number of rooms with a variable amount of reverberation to increase robustness with respect to separation performance in reverberant conditions.
While future datasets and studies should feature a higher degree of variability (number and kind of noise sources and speakers, head movements, moving sources, see below) we hope that open-sourcing our database fosters research of binaural separation algorithms for assistive listening devices in reverberant conditions. 

\textbf{Limitations of this work:}
The current study is constrained to static two-speaker scenes with diffuse noise with reverberation. 
This is a first step (and represents a typical communication scenario) for quantifying the general performance of our approach in comparison to a competitive baseline. 
In dynamically changing scenes, models trained with permutation-invariant training in \emph{static} scenarios could mix up signals from speakers, i.e., the output of the speaker could switch. 
To avoid such switching, our approach could be combined with location-based training as in \cite{location_based} or
online clustering of frame wise speaker embeddings as proposed in \cite{binaural_wave_split}. 

Kolbæk and colleagues have shown that a system with permutation-invariant training can be extended to more than two speakers (Kolbæk et al., 2017, \cite{upit}): A system trained to separate three speakers but using a two-speaker input produces a separation and one output with an energy that is 63\,dB below the other outputs. 
Since the proposed system also relies on permutation-invariant training, we assume that it can be extended to
three or more speakers as well (with the same constraints observed in this study regarding spatial separation angle and reverberation, and by increasing the size and complexity of the relatively large filter estimation module).

One aspect that was not explored in this study is related to usability, i.e., how to enable a selection of the desired speaker for the user of a hearing aid. 
One option is to choose the desired stream with an additional device (e.g., a smartphone). Alternatively, speaker separation in a listening device could be linked to directional cues and to electrooculography \cite{eye_gaze_eeg} (or optical eye tracking \cite{eye_gaze}) to determine the direction of eye gaze, thereby allowing a speaker selection. In \cite{invasiv_eeg} it was already shown that decoding the attended speaker can also be performed with electrical brain activity signals acquired form invasive electrodes.
Future hearing devices could also feature integrated or discrete behind-the-ear electrodes for decoding the attended speech stream from electrical brain activity \cite{eeg_decode_beam}. 

\textbf{Future work:}
In future work, it would be interesting to conduct a subjective evaluation of our models with hearing-impaired listeners, since subjective evaluation is still the gold standard for quantifying the effects of signal enhancement algorithms. 
From a technical point of view, a replacement of the fixed-size filter module with an alternative approach for predicting filters with a smaller number of parameters could be explored. 
To further reduce complexity, methods such as quantization \cite{fedorov20_interspeech} and pruning\cite{pruning} should be applied. The next step would be to evaluate the performance of such a model on an hearing-aid system-on-chip, potentially include head movements of the listener, and extend the model to more than two speakers. 
Another important future step is the further improvement of the audio quality, especially for high reverberation situations. This could be achieved through suppression of only the late part of the reverberation as performed in \cite{late_reverb} compared to the full dereverberation performed in this study. 
Alternatively, the output signal could be remixed with the unprocessed signal in a certain ratio \cite{remixing}, which increases awareness for the acoustic scene and can mask artifacts of the source separation procedure. 

\section{Conclusion}
This study introduced a simple binaural low-latency, low-complexity neural filter-and-sum beamforming model for speaker separation.
The model uses binaural input and uses relatively few parameters by using Group Communication and Transform Average Concatenate. 
We observe consistent improvements for four different metrics in challenging noisy conditions that cover different degrees of reverberation. 
With smaller models, performance drops gradually, while they still outperforming the larger baseline approaches in important metrics. 


\section{References Section}
\bibliographystyle{IEEEtran}
\bibliography{westhausen_23}

\vfill\pagebreak

\end{document}